\documentclass[reprint,aps,prl,amsmath]{revtex4-1}
\usepackage{setspace}
\usepackage{graphicx}
\usepackage[linkcolor=blue,citecolor=blue]{hyperref}
\usepackage{rotating}
\usepackage{float}
\usepackage{subfig}
\usepackage{amsmath,amssymb}
\usepackage{slashed}
\usepackage{bm}
\usepackage{mathrsfs}
\usepackage{tabto}
\hypersetup{colorlinks=true}
\usepackage[cal=boondoxo]{mathalfa}

\newcommand{\so}{{\bf S}}
\newcommand{\lo}{{\bf L}}
\begin{document}
\title{$P-$state Positronium for Precision Physics: an Ultrafine Splitting at $\alpha^6$}
\author{Henry Lamm}
\email{hlamm@umd.edu}
\affiliation{Department of Physics, University of Maryland, College Park, MD 20742}
\date{\today}

\begin{abstract}
An ``ultrafine'' splitting in positronium between the $L=1$ spin-singlet state and the spin-average of the spin-triplet states is shown to arise only at order $\alpha^6$.  The QED prediction for $n=2$ states is $\Delta_{2,P}=(683/172800)m\alpha^6=73.7(1.2)$ kHz.  This represents the smallest leading-order QED splitting known.  Current experimental efforts could observe this splitting, and its observation can constrain new ultralight interactions, such as axions or $Z'$.  
\end{abstract}
\maketitle
Spectroscopy of positronium represents a precision test of quantum electrodynamics (QED).  The low mass of the electron compared to other mass scales renders the contributions from strong, electroweak, and heavier lepton loops negligible within the accuracy of current experiments.  QED predictions for the positronium energy levels exist completely at order $\alpha^6$~\cite{Czarnecki:1999mw,Czarnecki:1998zv,Khriplovich:1993zz,Khriplovich:1993rh,PhysRevA.78.032103}. Research is active to compute the incomplete $\alpha^7$ correction~\cite{Karshenboim:1993,Melnikov:1999uf,Pachucki:1999zz,Kniehl:2000cx,Melnikov:2000zz,Hill:2000zy,marcu2011ultrasoft,Baker:2014sua,Adkins:2014dva,Eides:2014nga,Eides:2015nla,Adkins:2014xoa,Adkins:2015jia,Adkins:2015dna,Adkins:2016btu,Eides:2017uoy}.  These predictions are in agreement with existing measurements for a number of level transitions: $2^3S_1-1^3S_1$ ($1S-2S$ interval)~\cite{PhysRevLett.70.1397}, $1^3S_1-1^1S_0$ (ground state hyperfine splitting)~\cite{PhysRevA.30.1331,Ishida:2013waa}, and the $2^3S_1\rightarrow2^{3S+1}P_J$ (fine structure and Lamb shift)~\cite{Ley1994,PhysRevLett.71.2887}.

Energy levels are derived as a double power series in $\alpha^g\ln^h(\alpha)$ starting at $\alpha^2$.  The coefficients, $C_{gh}(n,L,J,S)$, can be zero (i.e. the $\alpha^3$ and $\alpha^4\ln^h(\alpha)$ terms).  Cancellations between $C_{gh}(n,L,J,S)$ can further occur in level transitions.  The $1S-2S$ interval depend only on spin-independent coefficients and begins at $\alpha^2$.  The fine structure and ground state hyperfine splittings (hfs) are sensitive only to spin-dependent terms and are nonzero at $\alpha^4$.  The Lamb shift (defined as the spin-independent $2S-2P$ splitting) starts only at $\alpha^5\ln(\alpha)$.
  
This work focuses on the hyperfine splitting between the spin-singlet state $n^1P_1$ and the spin-average of the spin-triplet $n^3P_J$ states:
\begin{align}
\label{eq:delta}
 \Delta_{n,P}\equiv& M(n^1P_1)\nonumber\\&-\frac{1}{9}\left[1\cdot M(n^3P_0)+3\cdot M(n^3P_1)+5\cdot M(n^3P_2)\right].
\end{align}
Naively, one might expect this splitting to be on the order of GHz like the Lamb shift or hfs.  Using the existing fine structure measurements (compiled in Tab.~\ref{tab:data}), one instead finds $\Delta_{2,P}=4.31(6.50)$ MHz~\cite{Mills:1975zz,Hatamian:1987zz,PhysRevLett.71.2887}. This result is consistent with zero within the experimental uncertainty at the parts per thousand level.  Analogous experimental splittings in heavy quarkonium are consistent with zero to even higher precision and Eq.~(\ref{eq:delta}) has been suggested as a method for studying the exotic spectrum of quarkonium~\cite{Lebed:2017yme}.

In this paper, we investigate the physical origin underpinning this precise cancellation in positronium.  We show that this particular hyperfine splitting arises only at $\alpha^6$ and is further suppressed parametrically, justifying calling it an \textit{ultrafine} splitting.  Finally, we briefly discuss how $\Delta_{n,P}$ constrains new ultralight interactions.  

\begin{table*}[t]
\begin{center}
 \caption{\label{tab:data}Experimental and theoretical values for the $n=2$ fine structure transitions.}
 \begin{tabular}{c c c c}
 \hline\hline
  Transition & $\Delta E_{exp} $(MHz)& $\Delta E_{theory}$(MHz) & $\delta 
E$(MHz)\\
\hline
$2^3S_1\rightarrow2^1P_1$ & 11180(5)(4)\cite{Ley1994} 
&11185.37(8)\cite{Czarnecki:1999mw}	&-5.37(500)(400)(8)	\\
$2^3S_1\rightarrow2^3P_0$ & 18499.65(120)(400)\cite{PhysRevLett.71.2887} 
&18498.25(8)\cite{Czarnecki:1999mw}	&1.40(120)(400)(8)	\\
 $2^3S_1\rightarrow2^3P_1$ & 13012.42(67)(154)\cite{PhysRevLett.71.2887} 
&13012.41(8)\cite{Czarnecki:1999mw}	&-0.01(67)(154)(8)	\\
 $2^3S_1\rightarrow2^3P_2$ & 8624.38(54)(140)\cite{PhysRevLett.71.2887} 
&8626.71(8)\cite{Czarnecki:1999mw}	&2.30(54)(140)(8)	\\
  \hline\hline
 \end{tabular}
\end{center}
\end{table*}

To understand the smallness of $\Delta_{n,P}$, it is useful to study the matrix elements of operators contributing to a $f\bar{f}$ bound state following the same discussion in the context of heavy quarkonium~\cite{Lebed:2017yme}.  The set of spin-independent operators depending on powers of squared momenta is infinite.  In stark contrast, only a finite set of linearly-independent spin-dependent operators exist given the fermion spins $\so_f,\so_{\bar{f}}$ and the orbital angular momentum $\lo$.  Wigner-Eckart theorem restricts nonzero matrix elements of spin operators in states with total spin $S$ to those which transform under an irreducible representation $k<2S$.  For states built from only two fermions, $S=0,1$ and therefore the largest irreducible representation is $k=2$.  

A particularly useful set of linearly-independent matrix elements up to quadratic order is  
\begin{align}
 \so_{f}\cdot\so_{\bar{f}}\phantom{xxx}&\text{(hyperfine)},\\
 \so\cdot\lo\phantom{xxx}&\text{(spin-orbit)},\\
 \tensor{T}\equiv(\so_{f}\cdot\hat{\bm{r}})(\so_{\bar{f}}\cdot\hat{\bm{r}})-\frac{1}{3}\so_{f}\cdot\so_{\bar{f}}\phantom{xxx}&\text{(tensor)},
\end{align}
where $\so\equiv\so_{f}+\so_{\bar{f}}$.  In a given spin-multiplet, all other spin-operators can be constructed from this set.   

Using Wigner-Eckart theorem,  the spin-orbit and tensor matrix elements must vanish for $S=0$ states because they transform as $S=1$ and $S=2$ respectively.  For $L=0$ states, these matrix elements are also zero, so any $n^3S_1-n^1S_0$ splitting depend only on $\so_{f}\cdot\so_{\bar{f}}$.  For states of $S=1$ and $L>0$, the spin-orbit and tensor matrix elements do not vanish but are given via the total angular momentum $J=L+1,L,$ and $L-1$:   
\begin{equation}
 \left<\so\cdot\lo\right>=\frac{1}{2}\left[J(J+1)-L(L+1)-S(S+1)\right]
\end{equation}
and 
\begin{equation}
\left<\tensor{T}\right>=\left\{\begin{array}{rl}
-\frac{L+1}{6(2L-1)} \, , & J = L-1 \, , \\
+\frac 1 6           \, , & J = L   \, , \\
-\frac{L}{6(2L+3)}   \, , & J = L+1 \, ,
\end{array} \right.
\end{equation}
In this multiplet, $(\so\cdot \lo)^2$ occurs as an order $\alpha^6$ correction and therefore the following relation is useful~\cite{landau2013quantum}:
\begin{align}
\label{eq:ls2}
 \left<(\so\cdot\lo)^2\right>=\frac{L(L+1)}{3}&\left<\so^2\right>-\left<\so\cdot\lo\right>\nonumber\\&-(2L-1)(2L-3)\left<\tensor{T}\right>,
\end{align}
where $\so^2$ has been used for compactness but is directed related to $\so_{f}\cdot\so_{\bar{f}}$.

Due to these additional non-zero matrix elements in $L>0$ states, individual $2^3S_1\rightarrow2^{3S+1}P_J$ transitions have a more complicated spin-dependence.  The spin-orbit and tensor elements can be made to vanish by construction in a sum weighted by the $2J+1$ degenerate spin states.  This sum receives contributions only from $\so_{f}\cdot\so_{\bar{f}}$ and is given by
\begin{align}
 \Delta_{n,L}\equiv& M(n^1L_{J=L})\nonumber\\&-\frac{2L-1}{3(2L+1)}M(n^3L_{J=L-1})\nonumber\\&-\frac{2L+1}{3(2L+1)}M(n^3L_{J=L})\nonumber\\&-\frac{2L+3}{3(2L+1)}M(n^3L_{J=L+1})
\end{align}
from which we have derived Eq.~(\ref{eq:delta}) for the $L=1$ case.  $\Delta_{n,L}$ can be though of as analogous to the hfs for $L>0$ states as both are purely hyperfine-operator dependent.  One might guess that $\Delta_{n,L}$ should be an order $\alpha^4$ observable like the hfs.  Instead, due to the dynamics underlying the non-relativistic hyperfine operator, $\Delta_{n,L}$ arises only at higher-order in positronium.  

The non-relativistic positronium potential $V(\bm{r})$ to order $\alpha^5$ has long been known (a modern derivation using NRQED can be found in ~\cite{Pineda:1998kn}).  The hyperfine operator arises at $\alpha^4$ and the only term up to $\alpha^5$ is proportional to $\delta^{(3)}(\bm{r})$~\cite{Fermi1930}, implying it arises as contact interaction from the reduction of the QED interaction.  A natural way to see this starts from the Dirac equation with the Breit interaction~\cite{PhysRev.39.616}.  In this Hamiltonian, the spin-spin coupling comes from the Laplacian operator acting on the Coulomb potential $1/r$, i.e. the Fourier transform of the massless propagator $1/q^2$. This term is a generic feature of massless gauge theories like QCD as well and therefore a source of hyperfine splitting in quarkonium. 

The non-relativistic wave functions near the origin scale as $r^{L}$, thus only $L=0$ states receive contributions from $\delta^{(3)}(\bm{r})$ terms.  Therefore $\Delta_{n,L}$ must come from at least $\alpha^6$ corrections with more non-trivial $r$-dependence.  The vanishing of $\Delta_{n,L}$ at $\mathcal{O}(\alpha^5)$ is a unique characteristic of positronium.  Heavier leptonic systems like true muonium ($\mu^+\mu^-$) has typical momentum $\alpha m_\mu\sim m_e$, and therefore receive corrections at $\alpha^5$ due to electron loop corrections to the potential~\cite{Jentschura:1997tv,Lamm:2017lib}.  For quarkonium states, both lighter flavors and the gluon self-coupling introduces corrections at $\alpha_s^5$~\cite{Lebed:2017yme}. 

For the $L=1$ states, the $\alpha^6$ contributions were computed in~\cite{Khriplovich:1993zz,Khriplovich:1993rh} (with small misprints corrected in \cite{Czarnecki:1999mw}).  The general $L>0$ state energy levels up to order $\alpha^6$ were computed in~\cite{PhysRevA.78.032103}.  From these, we find that the nonzero contributions to $\Delta_{n,P}$ come from
\begin{widetext}
\begin{equation}	
 \delta E=\frac{m\alpha^6}{n^3}\left[\frac{1}{2000}\left(46-\frac{43}{n^2}\right)(\so\cdot\lo)^2-\frac{1}{2000}\left(14-\frac{17}{n^2}\right)\so^2-\frac{1}{7680}\left(227+\frac{90}{n}-\frac{108}{n^2}\right)\kappa^2\right],
\end{equation}
\end{widetext}
where $\kappa=\frac{1}{5}\left(-2\left(\lo\cdot\so\right)^2+4\lo\cdot\so+\frac{4}{3}\so^2\right)$ and terms known to vanish in $\Delta_{n,P}$ have been neglected.  The remaining terms are all necessarily proportional to $\so_f\cdot\so_{\bar{f}}$.  This expression is the sum of a number of physically distinct processes: second-order relativistic corrections to the Coulomb potential, first-order relativistic corrections to single magnetic photon exchange, and iterating the Breit interaction\cite{Khriplovich:1993zz,Khriplovich:1993rh}.  These contributions are non-zero because they arise from higher moments of the wave function, $\left<\bm{r}^{-k}\right>$.

Using Eq.~(\ref{eq:ls2}), we can derive a general relation for reducing general spin-operator contributions into $\so_f\cdot\so_{\bar{f}}$ ones:
\begin{align}
 \Delta&_{n,L}\big((\so^2)^m(\lo\cdot \so)^n\big)=\nonumber\\&\frac{2^{m}}{9}\left[(-1-L)^n+3(-1)^n+5L^n\right]\Delta_{n,L}(\so_f\cdot\so_{\bar{f}}).
\end{align}
An expression for $\Delta_{n,L>1}$ could be derived by using the results of~\cite{PhysRevA.78.032103}, but we see that $\Delta_{n,L>1}\propto\alpha^6$ as well.  These splittings will be even more parametrically suppressed due to the larger $n$ required, but given that no experimental measurements have been undertaken to measure $L>1$ state energies, a non-zero measurement of $\Delta_{n,L>1}$ is not possible in the forseeable future.  For $P$-states, we then find that
\begin{equation}	
 \Delta_{n,P}=\frac{m\alpha ^6}{60} \left(\frac{137}{90 n^3}+\frac{1}{n^4}-\frac{1}{2n^5}\right)\Delta_{n,P}(\so_f\cdot\so_{\bar{f}}).
\end{equation}
using $\Delta_{n,P}(\so_f\cdot\so_{\bar{f}})=-1$ and setting $n=2$, we find
\begin{equation}
\label{eq:res}
 \Delta_{2,P}=\frac{683 m\alpha ^6 }{172800}=73.7(1.2)\text{ kHz}
\end{equation}
which is about $90$ times smaller than the existing experimental accuracy.  The smallness of $\Delta_{2,P}$ compared to other splittings is dramatic.  The hfs is 203 GHz, while the typical Lamb-shift-like splittings in Tab.~\ref{tab:data} are 10 GHz.  The theoretical error was estimated by considering the $\alpha$ expansion,  
\begin{equation}
 \Delta_{n,P}=m\alpha^6\left(C_0+C_{11}\alpha\ln\alpha+C_{10}\frac{\alpha}{\pi}+\hdots\right),
 \end{equation}
 where the coefficents $C_{gh}$ start at $\alpha^6$. Taking $C_0\sim C_{11}\sim C_{10}$, the $\mathcal{O}(\alpha\ln\alpha)$ term dominates the theoretical error in Eq.~(\ref{eq:res}).  The same estimate has was used to obtain the values in Tab.~\ref{tab:data}.  Comparing the results, we see removing the dependence upon $2^3S_1$ reduced the error by a factor of 60.

Having shown that existing measurements are too imprecise resolve this splitting, one can ask about the future prospects.  Upcoming experiments at University College London plan to remeasure the $n=2$ fine-structure with a reduced uncertainty yielding $\Delta_{2,P}\sim100$ kHz~\cite{Alonso:2015gvp,Alonso:2016faw,Alonso:2017qtd}.  A measurement of $\Delta_{2,P}$ would require only a factor of two improvement.  A precision of $\sim10$ kHz would translate into a $5\sigma$ detection.  

The spin-weighting of the ultrafine splitting means the uncertainty on $2^3S_1\rightarrow2^1P_1$ contributes 86\% of the error assuming the transitions are measured with equal accuracy.  The situation is worse because the $2^3S_1\rightarrow2^1P_1$ is a forbidden single-photon transition.  Previous measurements of this transition were performed in a magnetic fields where mixing with the other $2P$ states occurs.  With this mixing, the single-photon transition is allowed, and the multiple measurements at different magnetic fields are extrapolated to the zero-field limit via the known theoretical dependence~\cite{PhysRevA.8.625}.  This method inherently has larger uncertainties compared to the allowed $2^3S_1\rightarrow2^3P_J$ states.

Beyond measuring QED precisely, the ultrafine splitting can constrain new ultra-light spin-dependent interactions like the exchange of an axion or $Z'$.  The non-relativistic reduction of these interaction is a Yukawa potential $\alpha'\so_{f}\cdot\so_{\bar{f}}\frac{e^{-m_\phi r}}{r}$ where $\alpha'$ is the new coupling and $m_\phi$ is the mass of the new particle.  From this, the correction to $\Delta_{2,P}$ is
\begin{equation}
 \delta\Delta_{2,P}=\frac{m\alpha\alpha'}{8(1+2\frac{m_\phi}{m\alpha})^4}.
\end{equation}
Comparing this expression to Eq.~(\ref{eq:res}), we see that a measurement of $\Delta_{2,P}$ constrains $\alpha'\sim\alpha^5\sim10^{-11}$ in the $m_\phi=0$ limit.  This would be competitive with other model-independent constraints on ultralight particles in atomic systems~\cite{Karshenboim:2010cm,Karshenboim:2010cg,Karshenboim:2010cj,Karshenboim:2010ck,Karshenboim:2011dx,Lamm:2015lia}.

In this work, we have pointed out a previously overlooked ultrafine splitting in positronium.  At $\mathcal{O}(m\alpha^6)$, the $\Delta_{2,P}=0.0737(12)$ MHz splitting represents the smallest leading-order splitting from QED known.  While existing measurements are too imprecise, it should be resolved in the near-term.  Further theoretical work needs to be done to compute the $\mathcal{O}(m\alpha^7\ln\alpha)$ contribution.  Additionally, this very small effect presents a new opportunity to constrain ultralight particles with coupling to electrons.  

\begin{acknowledgments}
HL would like to thank R. Lebed for his suggestion of this topic and expert advice in its undertaking.  HL is supported by the U.S. Department of Energy under Contract No. DE-FG02-93ER-40762.
\end{acknowledgments}

\bibliography{/home/hlamm/wise}
\end{document}